\documentclass[journal = ancac3, manuscript = article, layout = twocolumn]{achemso}

\usepackage{graphicx}
\usepackage[]{xspace}

\def\wse2{WSe$_2$\xspace}
\def\rc{$\mu$RC\xspace}
\def\pl{$\mu$PL\xspace}
\def\ple{$\mu$PLE\xspace}
\def\ya{\textbf{y}-axis\xspace}

\title{Excitonic resonances in thin films of WSe$_2$: From monolayer to bulk material}
\author{Ashish Arora}
\email{ashish.arora@lncmi.cnrs.fr}
\affiliation{Laboratoire National des Champs Magn\'{e}tiques Intenses\\ CNRS-UJF-UPS-INSA, 25 rue des Martyrs, 38042 Grenoble, France}
\author{Maciej Koperski}
\affiliation{Laboratoire National des Champs Magn\'{e}tiques Intenses\\ CNRS-UJF-UPS-INSA, 25 rue des Martyrs, 38042 Grenoble, France}
\author{Karol Nogajewski}
\affiliation{Laboratoire National des Champs Magn\'{e}tiques Intenses\\ CNRS-UJF-UPS-INSA, 25 rue des Martyrs, 38042 Grenoble, France}
\author{Jacques Marcus}
\affiliation{Institut N\'{e}el, CNRS-UJF, BP 166, 38042, Grenoble, France}
\author{Cl\'{e}ment Faugeras}
\affiliation{Laboratoire National des Champs Magn\'{e}tiques Intenses\\ CNRS-UJF-UPS-INSA, 25 rue des Martyrs, 38042 Grenoble, France}
\author{Marek Potemski}
\email{marek.potemski@lncmi.cnrs.fr}
\affiliation{Laboratoire National des Champs Magn\'{e}tiques Intenses\\ CNRS-UJF-UPS-INSA, 25 rue des Martyrs, 38042 Grenoble, France}

\keywords{WSe$_2$, 2D semiconductors, excitons, trions}

\begin{document}
\begin{abstract}
We present optical spectroscopy (photoluminescence and
reflectance) studies of thin layers of the transition metal
dichalcogenide \wse2, with thickness ranging from mono- to
tetra-layer and in the bulk limit. The investigated spectra show
the evolution of excitonic resonances as a function of layer
thickness, due to changes in the band structure and, importantly,
due to modifications of the strength of Coulomb interaction as
well. The observed temperature-activated energy shift and
broadening of the fundamental direct exciton are well accounted
for by standard formalisms used for conventional semiconductors. A
large increase of the photoluminescence yield with temperature is
observed in WSe$_2$ monolayer, indicating the existence of
competing radiative channels. The observation of absorption-type
resonances due to both neutral and charged excitons in \wse2
monolayer is reported and the effect of the transfer of oscillator
strength from charged to neutral exciton upon increase of
temperature is demonstrated.
\end{abstract}


\section{\label{secIntro}Introduction}
Atomically thin layers of semiconducting transition metal
dichalcogenides (SC-TMDCs) of the form MX$_2$ (such as MoS$_2$,
MoSe$_2$, WS$_2$ and WSe$_2$) attract a considerable research
interest stimulated by the scientific curiosity to study a new
class of two-dimensional (2D) semiconductors and by their
potential applications in optoelectronics, photonics and the
development of
valleytronics.\cite{prl10km,nn12qhw,an13ge,an13szb,prl12dx,nn13amj,oe13pt}
Bulk MX$_2$ crystals, largely investigated in the
past,\cite{aip69jaw,jpc72arb,jpc76arb} consist of weakly bonded
X-M-X monolayer units (three atomic planes). Monolayer (1~ML) and
few-, N-monolayer (N~ML) flakes of SC-TMDCs can be conveniently
extracted from bulk crystals and deposited on substrates using the
methods of mechanical exfoliation.\cite{ksn05pnas} Since very
first works on MoS$_2$ layers\cite{prl10km}, it is now well
established that the band structure of N~MLs of SC-TMDCs
critically depends on N. Monolayers of SC-TMDCs are believed to be
direct band gap (or nearly direct band gap) semiconductors and
undergo a crossover to the indirect band gap structure of bulk
SC-TMDCs with increasing N.\cite{prb12wsy,acs13wz,prb12ar} Strong
excitonic (Coulomb) effects, whose strength may also vary with N,
is another relevant ingredient which determines the optical
properties of thin films of SC-TMDCs.\cite{prb12ar} A thorough
understanding of basic optical properties of these films is
essential for advancing their further explorations, towards
efficient future applications.

In this paper, we report on optical investigations of thin films
of \wse2 with thicknesses ranging from 1~ML to 4~ML and bulk
flakes, using micro-photoluminescence (\pl),
micro-photoluminescence excitation (\ple) and micro-reflectivity
contrast (\rc) spectroscopy techniques, in 5~K-300~K temperature
range. Our N-layer \wse2 structures are first characterized with
low temperature, band-edge \pl and absorption-type experiments
which depict the characteristic indirect to (nearly) direct band
gap crossover, when approaching from thicker to the 1~ML film. We
then focus on the evolution of absorption resonances as a function
of N in a broader energy range, as they are derived from \rc
measurements. The ground state A and spin-orbit (SO) split B
excitons associated with the K-point of the Brillouin zone are
each accompanied by the excited state resonances A$^*$ and B$^*$,
respectively. Energy separation between A and A$^*$, and between B
and B$^*$ varies with N, which reflects the changes in the
strength of excitonic binding as well as in the character of
excitonic states, from non-Rydberg states in 1~ML to rather
Rydberg like states in bulk. Another clear absorption resonance C,
observed in the upper spectral energy range and most sensitive to
N, is tentatively assigned to the transition at the M-point of the
Brillouin zone. Energy shift and broadening of the ground state A
exciton resonance with increasing temperature are both well
accounted for by conventional formulae for the band gap shrinkage
and optical phonon mediated broadening of free exciton emission,
which have been widely applied to other semiconductors.
Strikingly, the integrated PL intensity in the 1~ML of \wse2 is
found to increase considerably with temperature, whereas opposite
is observed in conventional direct band gap semiconductors.
Although neutral excitons (which dominate the PL spectra at high
temperature) have higher emission efficiency than the
bound/charged excitons (which dominate the PL at low
temperatures),\cite{prb14gw,arxiv14ia} this observation of an
enhancement of integrated intensity is a thinkable sign of the
presence of a radiatively inefficient channel at low temperature,
which upon thermalization, transfers the carriers to radiatively
efficient channel at higher temperatures. It opens up scope for
further investigations and discussions on this material. Finally,
both neutral (A) and charged (trion, T)\cite{prl93kk,prb95hb,nm13kfm,prl95gf} excitonic
resonances are observed in 1~ML flake not only in emission, but in
absorption-type \rc spectra as well. This allows us to demonstrate
the effect of transfer of oscillator strength from trion to
exciton, upon increasing the
temperature.\cite{prb00vc,jop03pk,pssb01ae,pssb01ras} Further
studies of the effects of the presence of free carriers on the
optical response of the WSe$_2$ monolayers are highly desirable.

\section{\label{secSamples}Samples and characterization}
\begin{figure}[!t]
\includegraphics[width=8cm]{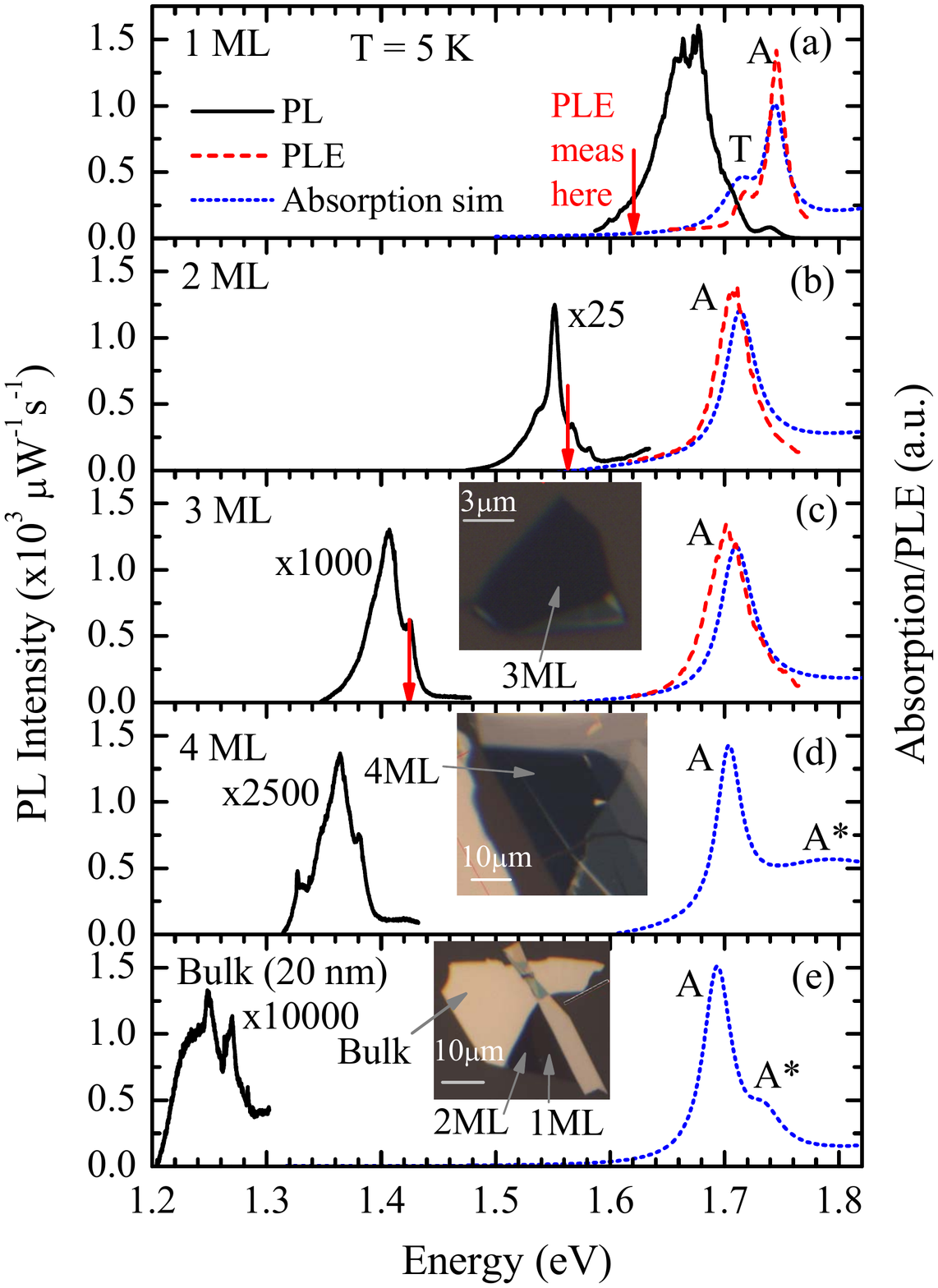}
\caption{Solid lines in (a-e) show the PL spectra for a 1~ML,
2~ML, 3~ML, 4~ML and 20~nm thick flakes respectively. Dashed red
lines in (a-c) are the PLE spectra of the corresponding flakes,
obtained by monitoring the PL emission intensity at an energy
shown by the arrows. The dotted curves are the absorption spectra
simulated from the modeling of the reflectance contrast
spectra of the corresponding flakes. The insets to (c) and (d)
show the optical image of a 3~ML and 4~ML thick flakes whereas the
inset to (e) shows a flake which contains 1~ML, 2~ML and 20~nm thick
bulk flakes used for the study.}\label{figPLPLE}
\end{figure}
For preliminary characterization of the flakes, we performed
atomic force microscopy (AFM) and \pl measurements on the flakes.
Insets to the Fig. \ref{figPLPLE} (c-e) show the optical images of
the flakes. The thicknesses of the 1~ML, 2~ML, 3~ML, 4~ML as well
as the bulk flake were found to be $\sim$0.7~nm, $\sim$1.4~nm,
$\sim$2.1~nm, $\sim$2.8~nm and $\sim$20~nm respectively using AFM
(data not shown). The PL spectra of the flakes have been shown as
solid curves in Fig. \ref{figPLPLE}(a-e). The 1~ML thick flake is
found to emit light most efficiently, and the emission intensity
reduces considerably as the thickness of the flakes is increased.
A reduction in intensity by 4 orders of magnitude is observed when
the flake thickness is increased from 1~ML to 20~nm. It is because
the nature of the band gap changes from direct to indirect for
thickness greater than 1~ML.\cite{acs13wz} The emission also
shifts to lower energy with increasing flake thickness. This
indicates the reduction of the fundamental bandgap with an
increased flake thickness, in analogy to the effect of quantum
confinement in semiconductor quantum wells.

To gain more insight into the near band-edge properties of our
layers, apart from \pl, we performed \ple and \rc spectroscopy
studies on the same flakes. The well pronounced, lowest energy
absorption resonance seen in these spectra (dashed blue and dashed
red traces in Fig. \ref{figPLPLE}) is identified with the
so-called A exciton associated to the direct band alignment of
\wse2 at K-point of the Brillouin zone. Its double peak structure
in the monolayer (trion T resonance on the low energy side) and in
the bulk sample (excited exciton state A$^*$ on the higher energy
side) will be discussed later on. A small shift in the absorption
resonances seen in the PLE spectra and those extracted from \rc
measurements is likely due to effectively higher sample
temperature in the course of PLE measurements which imply the use
of laser excitation and unavoidable local heating effects, and/or
due to energy sensitive carrier relaxation process which may also
alter the shape of absorption as reflected in PLE spectra.

One of the prominent and clear features of the data shown in Fig.
\ref{figPLPLE} is the red shift of the PL spectra with respect to
the fundamental absorption resonances. This shift progressively
increases with the number of layers. It is pretty large
($\sim$150~meV) for the \wse2 bilayer and up to $\sim$500~meV in
the bulk sample. The logic of indirect and low energy transitions
giving rise to the PL spectra but direct and higher energy
transitions determining the absorption resonances is consistent
with the indirect fundamental band gap in all \wse2 layers with
N$\ge$2. Instead, however, the monolayer \wse2 is believed to be a
direct band gap semiconductor. Its strong emission intensity and
partially overlapping emission and absorption spectra (practically
common emission and absorption peak due to A exciton) are in favor
of this scenario. The main part of the low temperature emission
spectrum of WSe$_2$ monolayer appears still below the fundamental
absorption resonance, what is, however, commonly attributed to the
recombination process due to localized/bound and/or charged
excitonic complexes.

\section{\label{secSamples}Absorption resonances versus layer thickness}
\begin{figure}[!t]
\includegraphics[width=8.5cm]{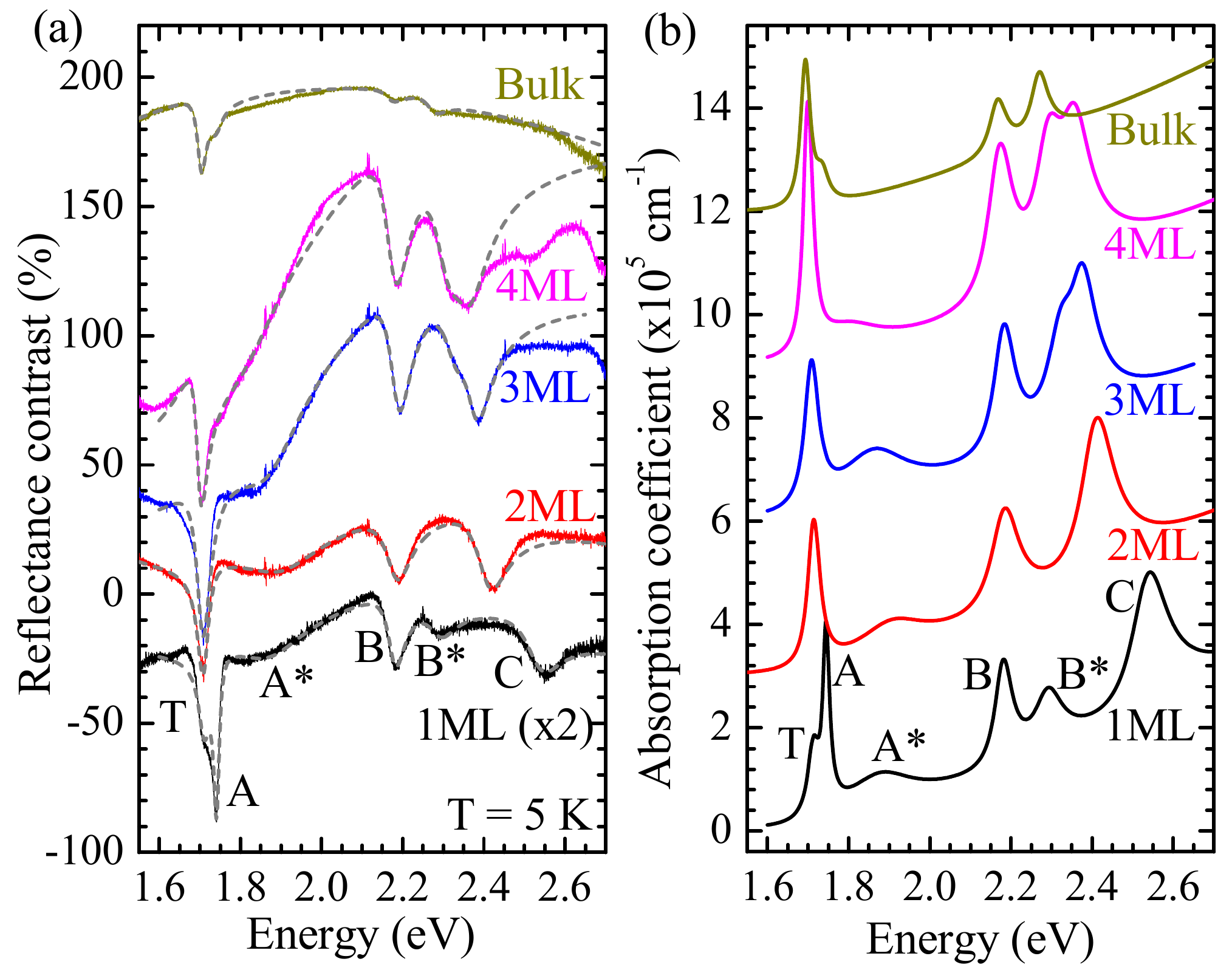}
\caption{(a) represents the reflectance contrast spectra of the
flakes shown in Fig. \ref{figPLPLE} with sample kept at 5~K
temperature. The dashed lines are the lineshape fits to the
corresponding spectra. (b) shows the absorption coefficient of the
flakes derived from the lineshape fittings. In both the plots, the
spectra for the layers with thickness $\ge$2~ML have been shifted
along the \ya for clarity.}\label{figLayerDepRef}
\end{figure}

In Fig. \ref{figLayerDepRef}(a), we present the results of RC
experiments performed at T~=~5~K for the same flakes in a wide
range of energy (spectra are shifted vertically for clarity).
Three dominant resonances, labelled A, B and C are observed for
all the different flakes. A and B are attributed to the two ground
state ($n=1$) excitonic absorptions at the K-point of the
Brillouin zone while the origin of the excitonic absorption C
(labeled as A$^\prime$ in some works following
Ref.~\citenum{aip69jaw}) is still a matter of
debate.\cite{aip69jaw,acs13wz,ns13wz,an14edc,arxiv14ak} Recent
\textbf{k$\cdot$p} calculations predict van-Hove singularities at
both K and M points, with a transition energy $\sim$3~eV for
1~ML.\cite{arxiv14ak} Considering an excitonic binding energy
equal to that of A-exciton (0.37~eV)\cite{prl14kh} as a first
approximation, then the exciton transition energy of $\sim$2.6~eV
agrees with our observation of C-exciton. Another report predicted
theoretically that the possible transition at M-point should red
shift by $\sim$100~meV when the flake thickness is increased from
2~ML to bulk.\cite{nl13wz} This is in qualitative agreement with
our observed red shift of $\sim$140~meV in C-exciton indicating
that this feature possibly originates from the M-point of the
Brillouin zone. In addition, the behaviour of C-exciton is in
contrast with that shown by the A-exciton transition where the
observed shift is only a $\sim$20~ meV, and the former is
therefore unlikely to have the much debated similar origins as
A-exciton i.e. at the K-point.  In all the flakes, A and B
excitons are accompanied by a broad shoulder at high energy side
of the excitons. We associate these features with the
contributions due to the excited state (n~=~2 and onwards) A$^*$
and B$^*$ exciton transitions. This is in accordance with a recent
study where the excited states of A exciton in \wse2 MLs were
observed at an energy range lying $\sim$160~meV to $\sim$370~meV
above the A resonance.\cite{prl14kh} A collective contribution of
all these states gives rise to a rather broad feature in RC
spectra. Furthermore, in the case of 1~ML thick flake, a feature
appears towards lower energy side of the A-exciton. As discussed
later, we identify this shoulder as the spectral contribution due
to the formation of the trion T (cherged exciton). Additional high
energy features appear in the RC spectra of the flakes thicker
than 2~ML. An assignment of these features requires a more
detailed theoretical and experimental work.

For the lineshape analysis of the RC spectra, we followed a method
similar to that described in Ref.~\citenum{jap13aa}. In this
method, we considered the excitonic contribution to the dielectric
response function to be given by a Lorentz oscillator like model
as
\begin{equation}
\varepsilon (E)=(n_b + ik_b)^2 + \sum_p{\frac{A_p}{(E_{p}^2-E^2-i\gamma_p E)}}.\label{eqnLor}
\end{equation}
where $n_b + ik_b$ represents the background complex refractive
index of \wse2 in the absence of excitons and was assumed to be
equal to that of the bulk material\cite{jpc76arb} in the
simulations. The index $p$ stands for the type of exciton
characterized by a resonance energy of $E_{p}$, with amplitude
$A_p$ and a phenomenological broadening parameter $\gamma_p$
[equal to full width at half maximum (FWHM) of the lorentzian
function]. The RC was then calculated using the transfer matrix
formalism. To fully reproduce our experimental data, we introduced
three additional oscillators in the dielectric function of the
\wse2 layers to account for the two observed exciton excited
states and for the trion feature. The refractive indices of Si and
the SiO$_2$ layers were obtained from Ref. \citenum{palik}. The
results of this calculation for our different flakes are presented
in Fig.~\ref{figLayerDepRef}(a) as dashed gray lines, together
with the corresponding experimental data. We notice that the
overall background shape of the reflectance contrast spectra has
been reproduced quite well using transfer matrix method. Also, the
Lorentz oscillator model has been able to reproduce the exciton
lineshapes nicely. It is then tempting to express these results in
the form of the absorption coefficient $\alpha(\lambda)$, defined
as
\begin{equation}
\alpha(\lambda)=\frac{4\pi k}{\lambda}.\label{eqnAbsSim}
\end{equation}
where $k$ is the imaginary part of the refractive index. The
calculated wavelength dependence of the absorption coefficient for
the flakes are presented in Fig.~\ref{figLayerDepRef}(b).

\begin{figure}[!t]
\includegraphics[width=7cm]{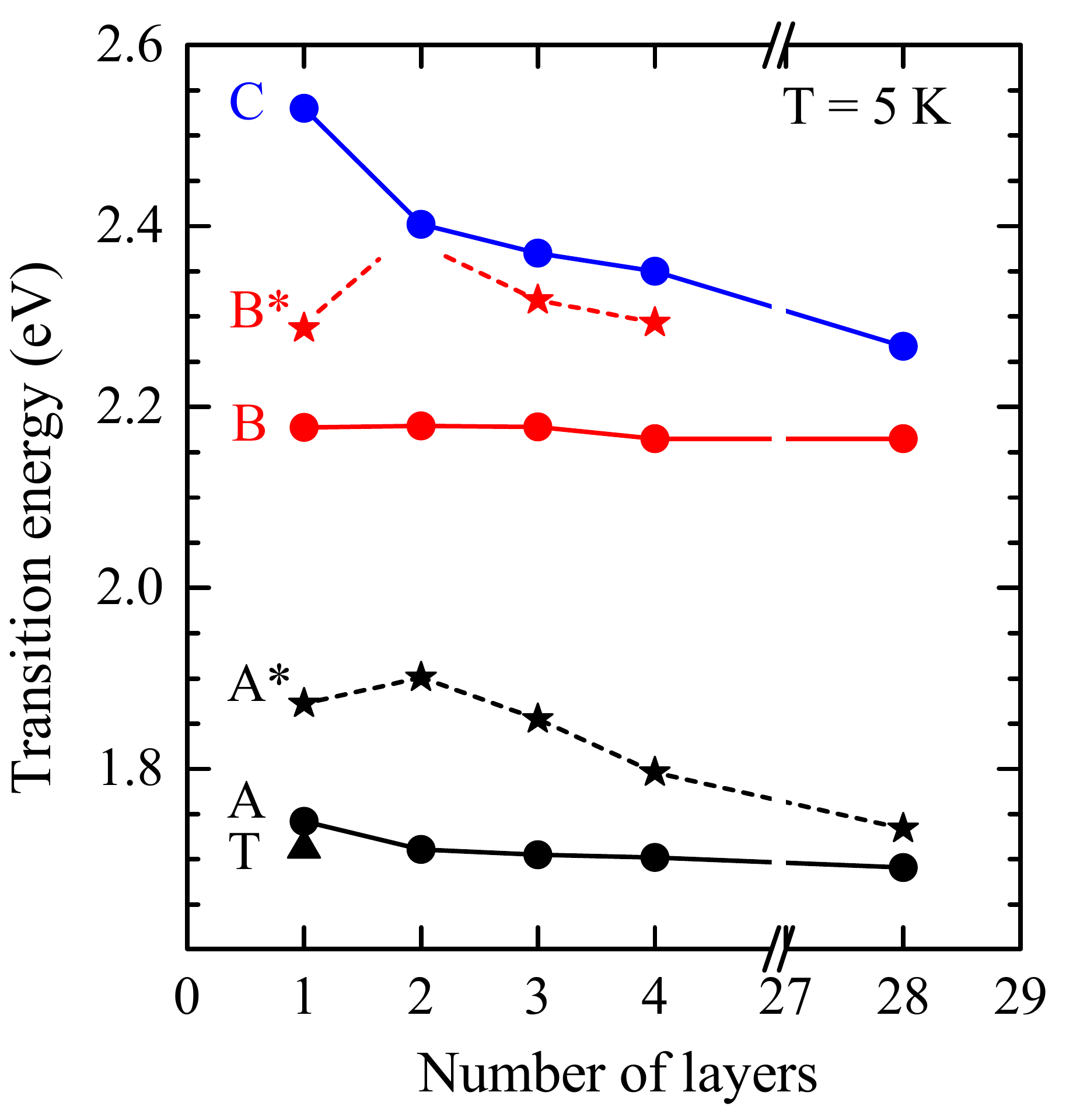}
\caption{Transition energies of the features obtained after modeling the reflectance
contrast spectra, for the excitons A, B and C, the excited state
contributions of the excitons A$^*$ and B$^*$ and the trion T as a
function of number of layers.}\label{figLayerDepEnergies}
\end{figure}

The transition energies derived from the modelling of our data are
presented in Fig. \ref{figLayerDepEnergies} as a function of the
flake thickness.
Both A and C-exciton energies decrease with increasing the number
of layers, the effect being most pronounced for C. The energy of
the B exciton appears to be only weakly dependent on the number of
layers. Interestingly, the energy of both excited states A$^*$ and
B$^*$ increases from  1~ML to 2~ML and then decreases for larger
thicknesses. This peculiar evolution observed only for the exciton
excited states is due to two competing effects. The first effect
is the reduction of the exciton binding energy, possibly due to
the change of the dielectric environment and of the dimensionality
of the material from 2D- towards a more 3D system. The second effect
is characteristic of TMDCs monolayers and arises from the nonlocal
character of screening of the Coulomb interactions in such
systems.\cite{prl13dyq,prl14kh,prl14ac} Indeed, exciton excited states in 1~ML
\wse2 strongly deviate from the hydrogenic Rydberg model,\cite{prl14kh} verified
for conventional 2D semiconductors. They were experimentally
observed much closer in energy to the ground state than
expected. This effect dominates in the case of the
1~ML flake but ceases to contribute when the number of layers is
increased. When going from the monolayer to the thicker flakes,
the excited state energy increases in a first step because the
exciton structure recovers its Rydberg character but
simultaneously, the exciton binding energy decreases when going
towards the 3D limit. The latter effect dominates for stacks
thicker than 3~ML and the excited state energy then decreases.
This scenario has to be confirmed on theoretical grounds.

Assuming that the dominant contribution to the A$^*$ structure for
the bulk comes from the n~=~2 state of the exciton, we find an A
exciton binding energy of $\sim$60~meV following the ideal Rydberg
model for 3D case. This is in agreement with a previous
report.\cite{jpc72arb} We notice that for 2~ML flake, the B$^*$
resonance merges with the C exciton peak and is omitted from the
fitting procedure. For the bulk flake, the B$^*$ structure is too
weak to be included in the analysis. From the RC fitting of the
1~ML thick flake, the trion (T) binding energy is estimated to be
$\sim$30~meV which agrees quite well with a recently reported
value.\cite{prb14gw}

\section{\label{secResults}Ground state exciton: Temperature activated energy shift and broadening}
\begin{figure}[!t]
\includegraphics[width=8.5cm]{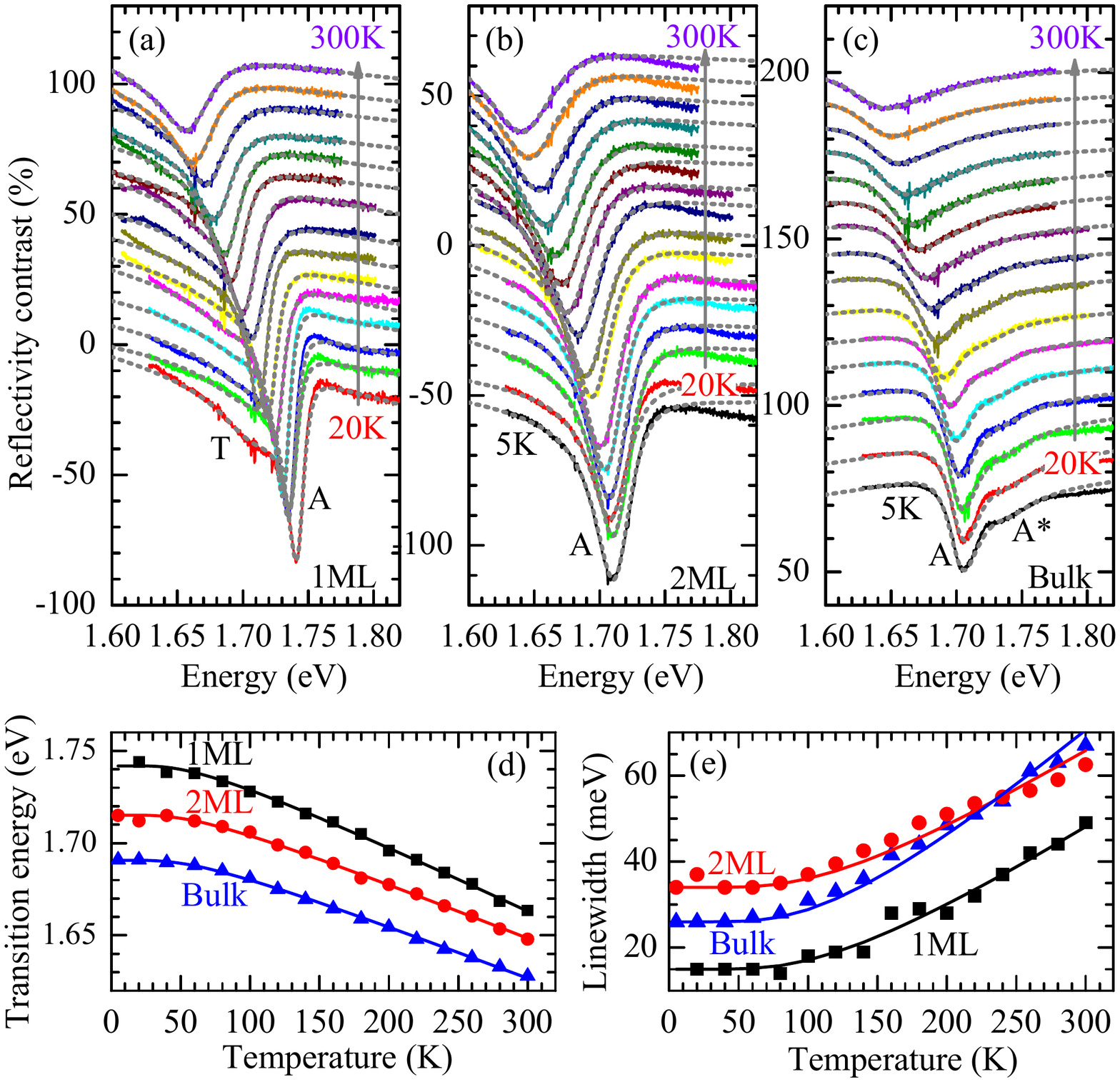}
\caption{(a-c) show the reflectance contrast spectra of the 1~ML,
2~ML and bulk flakes as a function of temperature, in an energy
region around the lowest energy exciton peak. The dashed lines
represent the corresponding lineshape fittings to the spectra.
The spectra were measured at 5~K (not available for 1~ML in this set of measurements), 20~K
and then upto 300~K in steps of 20~K. In all the cases, the
spectra have been shifted along the \ya for clarity. (d) Filled
squares, circles and triangles show the variation of the exciton
band gap of 1~ML, 2~ML and the bulk flake respectively, as a
function of temperature. The grey lines are the fits to the data
using Eq.~\ref{eqnODonnell}.}\label{figTempDepRef}
\end{figure}

To explore the properties of A exciton and T trions further, we
performed \rc spectroscopy of the 1~ML thick flake as a function
of temperature. In addition, we performed temperature dependent
\rc measurements of A exciton in a 2~ML and a 20~nm thick bulk
flake. Figures~\ref{figTempDepRef}(a-c) show the evolution of the
RC spectra around the A exciton peak as a function of temperature
(5~K to 300~K) for 1~ML, 2~ML and 20~nm thick flake respectively,
along with the calculated spectra. In the fitting procedure, we
have also considered T for 1~ML and A$^*$ for the bulk flake.
Commonly for the three studied films, the exciton resonance
energies undergo a red shift with increasing temperature as shown
in Fig.~\ref{figTempDepRef}(d). This evolution, very typical for
semiconductors, can be reproduced by the bandgap versus
temperature relations, equally well when using the
Varshni\cite{physica67ypv} or O'Donnell \textit{et
al.}\cite{apl91kpo} formulae. Varshni's relation is given by:
\begin{equation}
E_g(T)=E_0-(\alpha T^2)/(T+\beta).\label{eqnVarshni}
\end{equation}
where $E_0$ is the band gap at absolute zero, $\alpha$ and $\beta$
are the fitting parameters related to the temperature-dependent
dilatation of the lattice and Debye temperature respectively.
Although, this relationship describes the evolution of band gap,
it reproduces the exciton resonance energies equally well, which
suggests the independence of exciton binding energy on
temperature.
\begin{table}
  \caption{Fitting parameters as obtained from fitting of temperature dependence of the ground state exciton resonance energies and lineshape broadenings for 1~ML, 2~ML and bulk \wse2 flakes}
  \label{tableVarshni}
  \begin{tabular}{lccc}
    \hline
    Parameter  & 1~ML & 2~ML & Bulk  \\
    \hline
   Using Equation \ref{eqnVarshni} & \\
    \hline
    $E_0$ (eV)  & 1.744 & 1.717 & 1.692 \\
    $\alpha (\times 10^{-4} eV K^{-1})$  & 4.24 & 3.58 & 3.44 \\
    $\beta$ (K)  & 170 & 170 & 170 \\
    \hline
    Using Equation \ref{eqnODonnell} & \\
    \hline
    $E_0$ (eV)  & 1.742 & 1.716 & 1.691 \\
    $\langle \hbar\omega \rangle$ (meV)  & 15 & 15 & 15 \\
    S  & 2.06 & 1.75 & 1.67 \\
    \hline
    Using Equation \ref{eqnSegall} & \\
    \hline
    $\gamma_0$ (meV)  & 15 & 34 & 26 \\
    $\gamma^\prime$ (meV)  & 78 & 75 & 105 \\
    \hline
  \end{tabular}
\end{table}
The parameters which fit the data are given in
Table~\ref{tableVarshni}. We notice that the parameter $\alpha$
shows a reduction as the thickness of the flake is increased from
1~ML to 2~ML and bulk. Notably, there is a fundamental change in
the crystal lattice when the thickness is increased from 1~ML
thickness onwards. However, the $\beta$ parameter is $\sim$170~K
for all the three kinds of flakes, and it was fixed to this value
during the fitting. The effective Debye temperature seems
therefore to be independent of the film thickness.
Our estimation of
$\beta$ parameter of \wse2 is in agreement with a previous report
of Hu et al. where a value of $\beta=160$~K was found from the
photoconductivity measurements on bulk crystals.\cite{pssa07hu}
The relation provided by O'Donnell et al. describes the band gap
dependence on temperature in terms of average phonon energy
$\langle \hbar\omega \rangle$, as the following
\begin{equation}
E_g(T)=E_0-S\langle \hbar\omega \rangle[\coth(\langle \hbar\omega \rangle/2kT)-1].\label{eqnODonnell}
\end{equation}
where S is a coupling constant. The parameters obtained for the
best fit are given in Table \ref{tableVarshni}. Similar to $\beta$
parameter, $\langle \hbar\omega \rangle$ stays nearly the same
$\sim$15~meV, and its value was fixed during fitting. The S
parameter shows a gradual decrease from the monolayer to the bulk.
Both relations produce seemingly indistinguishable fits to the
data, and the curves corresponding to Eq.~\ref{eqnODonnell} are
shown in Fig.~\ref{figTempDepRef}(d).

Figure~\ref{figTempDepRef}(e) shows the temperature dependence of
linewidth broadenings ($\gamma_p$ in Eq.~\ref{eqnLor}) obtained
from the fits in Figs.~\ref{figTempDepRef}(a-c). In
semiconductors, the temperature dependence of the width of the
n~=~1 ground state exciton has the form\cite{prb90sr}
\begin{equation}
\gamma(T)=\gamma_0+\sigma T+\gamma^\prime \frac{1}{e^{\hbar\omega/kT}-1}.\label{eqnSegall}
\end{equation}
where $\gamma_0$ is the broadening at 0~K, the term linear in T
depicts the interaction of excitons with acoustic phonons and the
last term arises from interaction with LO (longitudinal optical)
phonons. The former is neglected being negligibly small compared
to the latter, which is proportional to the Bose function for
LO-phonon occupation. $\hbar\omega$ is the LO-phonon energy and is
taken to be equal to 31.25~meV (250~cm$^{-1}$) for
\wse2.\cite{oe13pt} The fits to the data are shown as solid lines
in Fig.~\ref{figTempDepRef}(e) and the fitting parameters are
shown in Table~\ref{tableVarshni}. We notice that $\gamma_0$,
which depends on the film quality, is different for the three
films studied. However, $\gamma^\prime$ is similar for 1~ML and
2~ML flakes, whereas it is larger for the bulk film. It can be
understood as follows. For 1~ML and 2~ML films, the binding energy
of the exciton is much larger than the LO phonon energy, therefore
the probability of scattering of the ground state exciton to its
excited bound and continuum states through the LO-phonons is
negligible.\cite{prb90sr} In such a case, the scattering of the
exciton assisted by an annihilation of LO-phonons is mostly
permissible within the n~=~1 state, to higher in energy excitonic
states with larger center-of mass momentum. However, for the
bulk case, the exciton binding energy is comparable to that of
LO-phonon energy. Therefore, the ground state exciton has a
significant probability to scatter to the energetically accessible
excited exciton states through annihilation of an LO-phonon. Such
additional contributions give rise to an increase in broadening in
the bulk flake as compared to the 1 and 2~ML flakes at elevated
temperatures, in turn increasing the slope of the $\gamma$(T)
curve.

\section{\label{Results}Excitons and charged excitons in 1~ML flake}
\begin{figure}[!t]
\includegraphics[width=8.5cm]{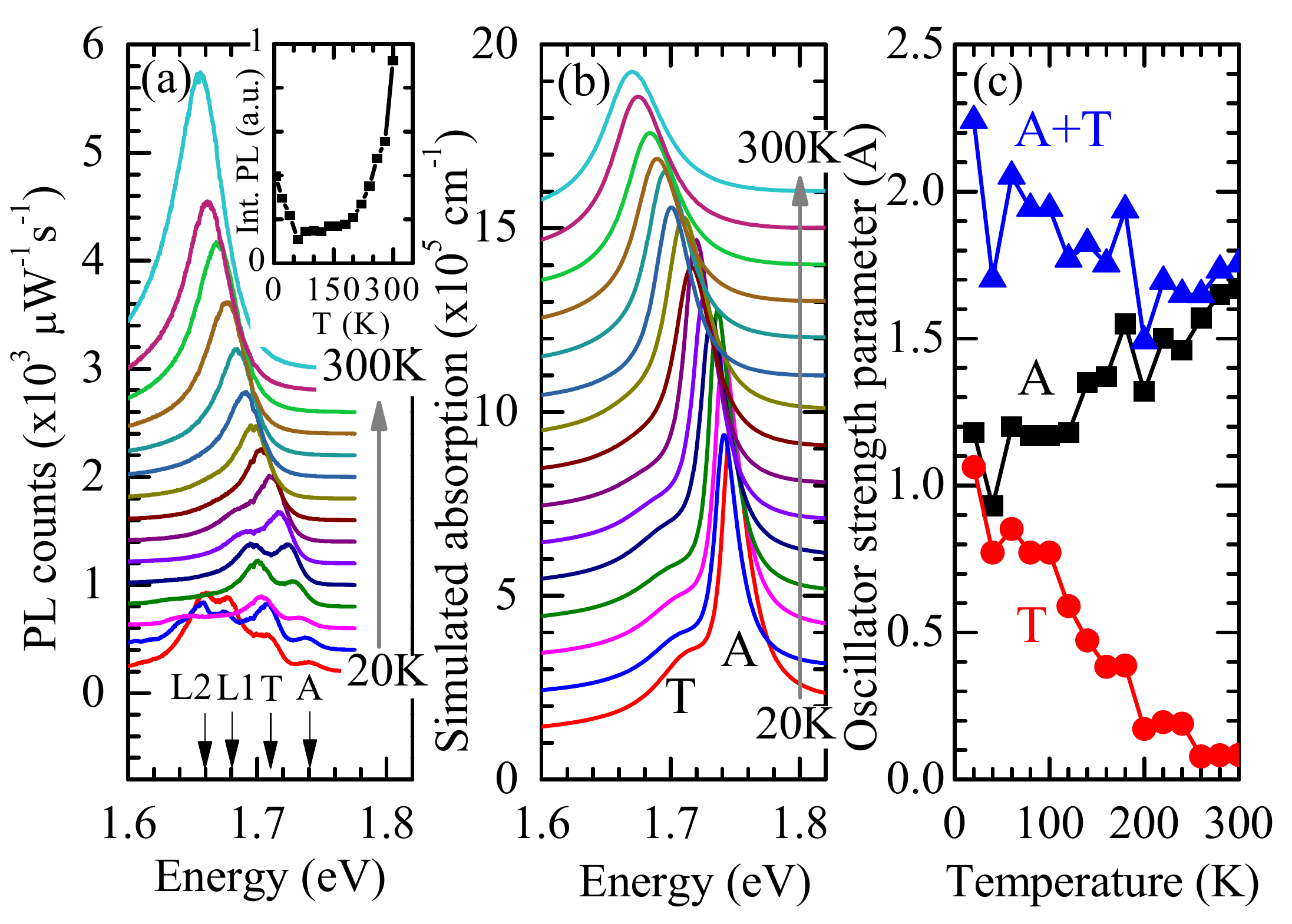}
\caption{(a) and (b) show the PL and absorption spectra (derived
from reflectance) for 1~ML thick flake as a function of
temperature, in an energy region around the lowest energy exciton
peak. The arrows in (a) represent the energy positions of the
indicated spectral features at 20K and the inset shows the
integrated PL intensity as a function of temperature. (c) Squares
and circles show the temperature dependence of the oscillator
strength parameter of the exciton, A and the trion, T. Triangles
show the total oscillator strength
parameter.}\label{figTrionsAndExc}
\end{figure}

The monolayers of SC-TMDCs are commonly accepted to be the direct band gap semiconductors and their
optical response in the vicinity of the band edge is understood in terms of the associated excitonic
resonances (free-, charged- and localized/bound excitons). At low temperatures and in the presence of
impurities and/or free carries, the localized/bound and/or charged exciton states (trions) usually
dominate the PL response of direct gap semiconductors whereas the free exciton emission is rather
weakly pronounced. This picture can also be applied to account for low temperature spectra of our
\wse2 monolayer, as shown in the bottom traces in Fig.~\ref{figTrionsAndExc}(a). These spectra
consist of up to four emission peaks which, in line with the previous studies, can be attributed
to the neutral exciton (A), charged exciton (T, trion) and two localized/bound excitons
(L1 and L2).\cite{prb14gw} The observed evolution of the \pl spectra with temperature is,
however, intriguing. To begin with, when the temperature is raised, the L1
and L2 peaks disappear from the spectra (at $\sim$40~K) initially whereas the higher energy
trion emission survives up to considerable higher temperatures (up to $\sim$200~K). The
opposite would be expected when naively assuming that lower in energy excitonic complexes
should have higher dissociation/activation energies. This assumption may not, however, apply
to TMDCs' monolayers for which the T complexes have been speculated to exhibit higher
dissociation energies than the impurity-bound excitons\cite{arxiv14ia}, in overall agreement
with our observation. Regardless of the apparent dissociation energies, the PL dynamics
is another relevant parameter which has to be taken account when discussing the PL changes
with temperature. Time resolved PL studies of \wse2 monolayers show that
A-exciton emission decays rather fast (few picoseconds decay time) whereas the decay
time of the subsequent lower in energy PL peaks progressively increases up to
$\sim$100~ps for the L2 peak.\cite{prb14gw} The temperature-activated population of higher
energy states, which presumably give rise to the emission more effectively could also
qualitatively account for the observed temperature-activated redistribution of
the PL intensity among different emission peaks.

This redistribution itself can be qualitatively understood but the observed rise of
the integrated PL intensity with temperature (inset to
Fig.~\ref{figTrionsAndExc}(a)) is a veritable puzzle. Typically in semiconductors,
a quenching of the PL intensity is observed with increasing temperature.
The observed rise of the PL yield (integrated intensity) is 
particularly striking because it appears at relatively higher temperatures 
where all the emission appears to arise due
to recombination of free excitons. We do not have a solid explanation for 
this effect but cannot restrain from speculating on its possible origin. We put forward
an hypothesis where we speculate an 
existence of certain dark, optically inactive states in the \wse2 monolayer, 
below the ground state A exciton. Those dark 
states could be associated with the excitons indirect in k-space, which is
however in contradiction with a general belief that the \wse2 monolayer is 
a direct gap semiconductor,\cite{acs13wz} (though the consensus in this matter 
may not have been reached so far\cite{prb12ar}). Optionally one could speculate 
about other dark excitonic states, similar to the exchange-split dark excitons 
in conventional semiconductors,\cite{prb88yc} in case of the specific 
alignment of the signs of the spin-orbit splitting in the conduction and 
valence bands of the \wse2 monolayer.\cite{prb13kk} The existence of such 
states could also qualitatively account for a double exponential decay of 
the A-emission reported recently from time resolved experiments.\cite{prb14gw}
Nonetheless, firm conclusions about the fine structure of the fundamental 
exciton in \wse2 monolayer require further experiments and theoretical studies.

The appearance of charged excitons in SC-TMDCs has been largely discussed in 
literature,\cite{nn13amj,nm13kfm} though their identification has been so far 
based on the PL spectroscopy only. Most convincing with this respect are the 
PL experiments carried out as a function of the gate voltage 
(carrier concentration),\cite{nn13amj} nevertheless the distinction between 
the bound/localized and (free) trion states as observed in low temperature 
PL is rather a subtle issue. Vast studies of conventional two dimensional 
structures (e.g. CdTe\cite{prl93kk} and GaAs\cite{prl95gf}) show that 
charged excitons are characterized by the significant oscillator strength and 
should be also visible in absorption type spectra, whereas bound/localized 
excitons may be predominant in the PL spectra. With the results shown in 
Fig.~\ref{figTrionsAndExc}, obtained for one of our 1~ML layer flakes, we 
identified the trion peak (T) on the low energy side of the A exciton in the 
low temperature PL [Fig.~\ref{figTrionsAndExc}(a)] and in the absorption type 
spectra [Fig.~\ref{figTrionsAndExc}(b)] as well. Observation of trions in 
absorption spectra is further confirmed by spectral evolution with temperature. 
A characteristic transfer of the oscillator strength from the trion to neutral 
exciton is observed upon increasing the temperature. This effect is analogous to 
what has been largely discussed in the context of optical studies of 
(weakly doped) CdTe and GaAs quantum wells.\cite{prl95gf,prb00vc,jop03pk,pssb01ae,pssb01ras} 
Generally speaking, the loss and gain of the oscillator 
strength of charged and neutral exciton with temperature, respectively, is due to the 
thermal redistribution of free carriers from Fermi-Dirac statistics at low 
temperature towards Boltzmann distribution at higher temperatures. The increase 
of temperature opens the k-space around the band edge for the efficient formation of 
the neutral exciton which leads to its increased oscillator strength. On the other 
hand, the weakening of the oscillator strength of charged excitons associated 
with carriers with higher k-wave vectors explains the progressive disappearance 
of the trion absorption with temperature. Quantitative analysis of the transfer 
of the oscillator strength from trion to neutral exciton remains to be done in future, 
on samples with the controlled carrier concentration. Interestingly,
the evolution of the optical response with carrier concentration may scale very 
differently in SC-TMDCs monolayers than in conventional semiconductors, due 
specific form of Coulomb interaction and non-Rydberg character of excitons 
in the former systems.\cite{prl14kh,prl14ac}

\section{\label{Conclusions}Conclusions}
In conclusion, we have studied the excitonic band structure of
layered \wse2 semiconductor as a function of layer thickness and
temperature. We presented a comprehensive lineshape analysis of
the RC spectra, and deduced the exciton transition energies. The
exciton transitions taking place at the K-point and M-point of the
Brillouin zone were observed in the spectral energy range of
$\sim$1.6~-~2.7~eV. The parameters dictating the temperature
activated energy shift and broadening of the fundamental
absorption resonance have been determined for 1~ML, 2~ML and bulk
\wse2 samples. The formation of a charged exciton (trion) in \wse2
monolayer has been traced with absorption type experiments and the
transfer of the oscillator strength from the trion to exciton has
been observed upon increasing the temperature.  The
observation of a striking rise of the PL yield with temperature in
a monolayer \wse2 opens up a scope for further theoretical and
experimental investigations of this system.

\section{\label{Methods}Methods}
The monolayer (ML) and a few layer (FL) \wse2 flakes were obtained on Si/(100~nm)SiO$_2$ substrate using mechanical exfoliation techniques.\cite{ksn05pnas} The flakes of interest were first identified by visual inspection under a microscope. A transfer matrix based calculation was performed to predict the color of the flake as a function of its thickness.\cite{hecht} The observed color of the flake was then compared with the theoretically predicted one to roughly estimate the number of layers in the flakes. For performing \pl measurements, the 632.8~nm radiation from He-Ne laser was focused on the flake using a 50x long working distance objective. The sample was mounted on the cold finger of a continuous flow liquid helium cryostat, at a temperature of $\sim$5~K. The spot diameter was $\sim$2~$\mu$m and the light power focused on the sample was 5~$\mu$W for 1~ML, 50~$\mu$W for 2~ML and 250~$\mu$W for the thicker layers. The PL emission from the sample was dispersed using a 0.5~m focal length monochromator and detected using a liquid nitrogen cooled Si charge coupled device camera. The setup for performing \ple was similar to the one used for \pl, except that a tunable Ti-sapphire laser with wavelength ranging from 700~nm to 800~nm was used for excitation. The intensity of the emission at a certain detection wavelength for each flake was monitored as a function of the excitation wavelength to record the PLE spectrum. For performing the \rc measurements, the light from a 100~W tungsten halogen lamp was focused on a pinhole of 150~$\mu$m diameter. The light was then collimated and focused (spot size~$\sim$4$~\mu$m) on the sample. The reflected light was detected using a setup similar to the one used for performing \pl. If $\mathcal{R}(\lambda)$ and $\mathcal{R}_0(\lambda)$ are the reflectance spectra of the \wse2 flake and the Si/SiO$_2$ substrate respectively, as a function of the wavelength ($\lambda$), then the percentage reflectance contrast spectrum is defined as follows:
\begin{equation}
\mathcal{C}(\lambda)=\frac{\mathcal{R}(\lambda)-\mathcal{R}_0(\lambda)}{\mathcal{R}(\lambda)+\mathcal{R}_0(\lambda)}.\label{eqnRefContrast}
\end{equation}
The measurements were performed at temperatures ranging from 5~K to 300~K. It must be mentioned that the RC spectrum shown for 1~ML flake in Fig. \ref{figLayerDepRef} was measured after approximately 1~month than the spectra shown in Fig. \ref{figTempDepRef}(a). We notice that the overall spectral intensity of the RC exciton features was reduced to about half when the sample was around 1~month old due to the effects of sample aging. In addition, the oscillator strength of the trion peak showed an increase after 1~month, which is possibly due the spontaneous doping of the sample in atmosphere. However, the energy positions of the spectral signatures did not show any noticeable shift.

\begin{acknowledgement}
We thank Piotr Kossacki for useful discussions, Ivan Breslavetz for technical support and Sandip Ghosh for valuable suggestions during building of the \rc setup. We acknowledge the support from the EC Graphene Flagship project (No. 604391) and the European Research Council (MOMB project No. 320590).
\end{acknowledgement}

\end{document}